\documentclass[]{elsart}
\usepackage{graphics}
\usepackage{graphicx}
\usepackage{epsfig}
\usepackage{amssymb}
\begin{document}
\begin{frontmatter}
\title{Traders' strategy with price feedbacks in financial market}


\author{\small{Takayuki Mizuno$^1$$^{a}$}},
\author{\small{Tohur Nakano$^{a}$}},
\author{\small{Misako Takayasu$^{b}$}},
\author{\small{Hideki Takayasu$^{c}$}}

\address{$^{a}$Department of Physics, Faculty of Science and Engineering, Chuo University, 1-13-27 Kasuga, Bunkyo-ku, Tokyo 112-8551, Japan}
\address{$^{b}$Department of Complex Systems, Future University-Hakodate, 116-2 Kameda-Nakano-cho, Hakodate, Hokkaido 041-8655, Japan}
\address{$^{c}$Sony Computer Science Laboratories Inc., 3-14-13 Higashigotanda, Shinagawa-ku, Tokyo 141-0022, Japan}

\thanks{Corresponding author.\\
{\it E-mail address:}\/ mizuno@phys.chuo-u.ac.jp (T.Mizuno)}

\begin{abstract}
We introduce an autoregressive-type model of prices in financial market 
taking into account the self-modulation effect. We find that traders are 
mainly using strategies with weighted feedbacks of past prices. These 
feedbacks are responsible for the slow diffusion in short times, apparent 
trends and power law distribution of price changes.
\end{abstract}

\begin{keyword}
Econophysics \sep Financial market \sep Self-modulation effect
\PACS 89.65.Gh,02.50,05.40.+j
\end{keyword}

\end{frontmatter}

\textbf{Introduction} \\
Very short time scale behaviors of yen-dollar market prices show 
significant deviation from a pure random walk such as slower diffusion [1] 
and non-trivial up-down statistics [2]. These properties are expected to be 
directly caused by traders' strategy. 

Recently, the analysis of trade intervals has revealed a peculiarity of 
traders' action that the occurrence of transactions can be nicely modeled by 
the self-modulation process with characteristic time about 2 minutes [3,4]. 
The self-modulation process is a stochastic process of which basic 
parameters such as the mean value are modulated by the moving average of its 
own traces, and it is proven rigorously that the power spectrum of resulting 
time series is inversely proportional to the frequency called the $1 / 
f$ spectrum [5]. In this paper we apply the idea of self-modulation to the 
analysis of price fluctuations and derive a new model of market prices.

\begin{center}
\begin{figure}
\begin{minipage}{.5\linewidth}
\centerline{\includegraphics[height=43mm]{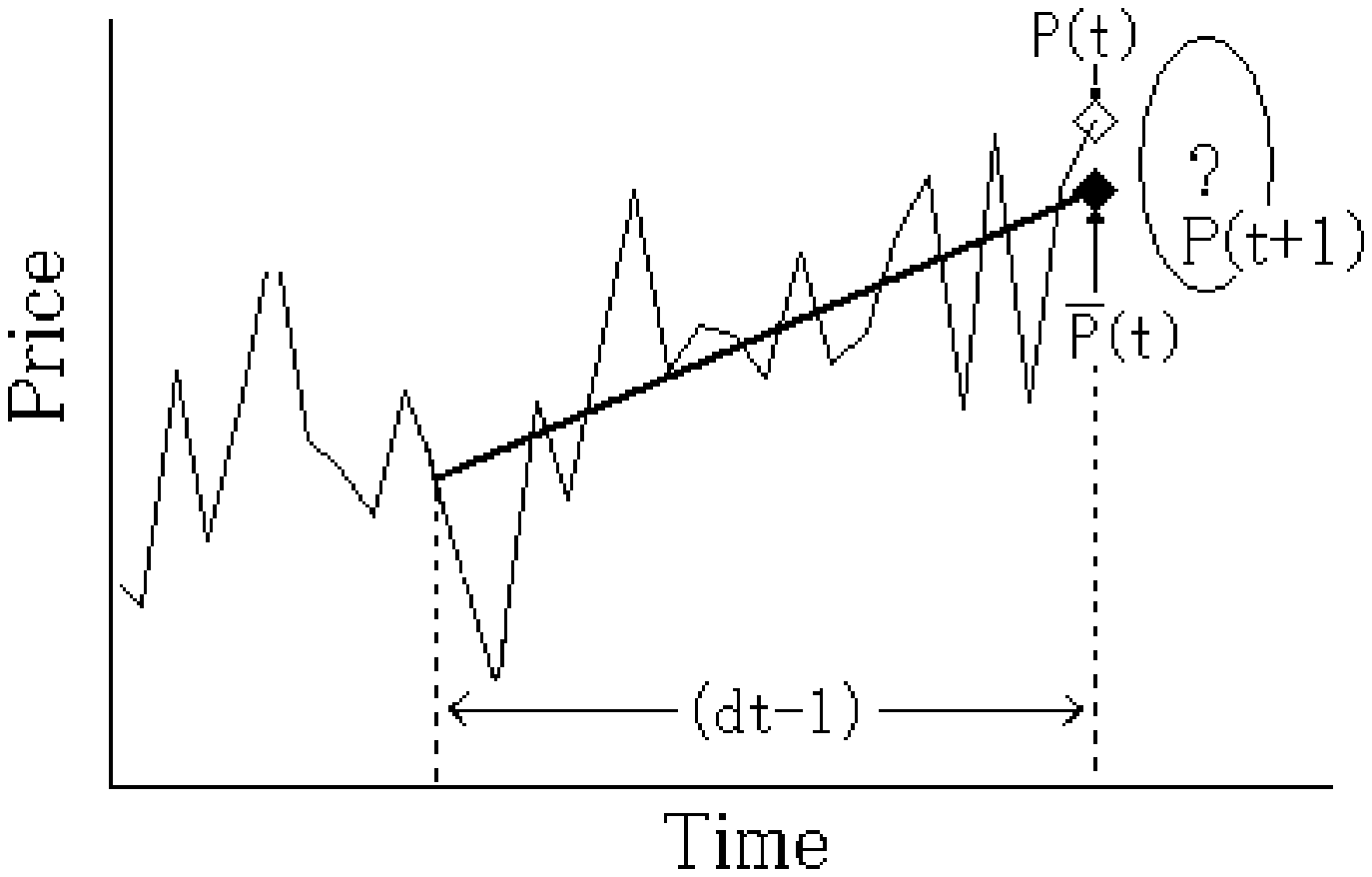}}
\label{fig1}
\caption[Fig.1 ]{Basic strategy of traders. Traders predict the mean market price 
$\overline P (t)$ from past prices between time $t - dt + 1$ and time $t$. 
The next quote, $P(t + 1)$, is expected to be in the circle in this figure.}
\end{minipage}
\begin{minipage}{.5\linewidth}
\centerline{\includegraphics[height=43mm]{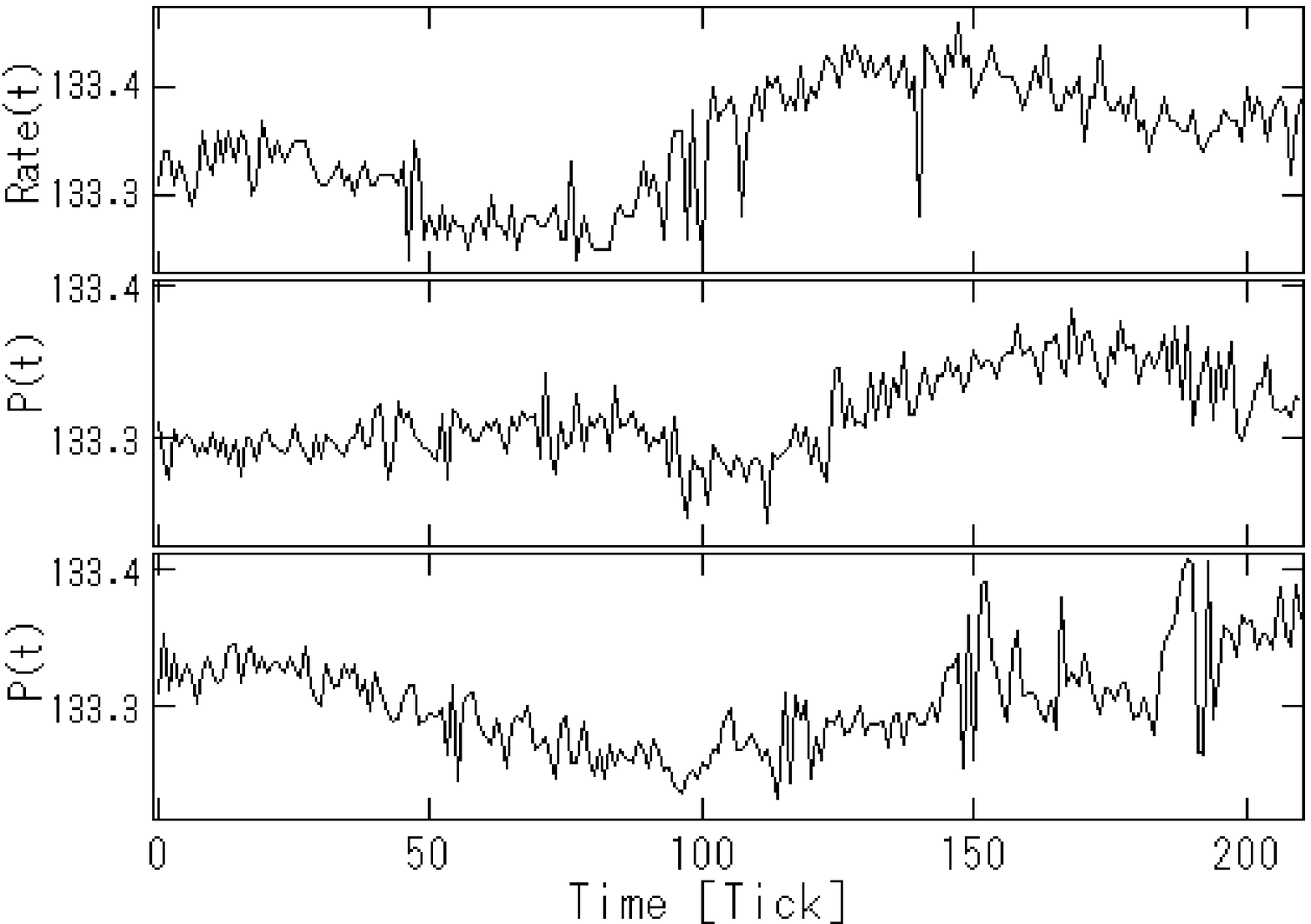}}
\label{fig2}
\caption[Fig.2 ]{An example of prices of (Top) actual market and 
(Middle and Bottom) those of 
simulation results using the real data for $t < 0$. 
Parameters are $dt = 30$, $b_1 = 0$ and $C = 1.7$. 
Random numbers of bottom differ from that of middle.}
\end{minipage}
\end{figure}
\end{center}

\vspace{-10mm}
\textbf{Traders' basic strategy model} \\
We introduce a kind of autoregressive model of market price that the mean 
market price $\overline P $(t) is given by a weighted moving average of past 
market prices {\{}$P(t)${\}}. We assume that a trader at time t observes a 
raise-trend of price as shown in Fig.1, then he may quote a price $P(t + 1)$ 
at time $t + 1$ in a circle drawn in Fig.1. The mean market price $\overline 
P (t)$ and a market trend $\Delta \overline P (t)$ at time $t$ are given by 
the following equations,

\begin{equation}
\label{eq1}
\overline P (t) = \frac{1}{2}(dt - 1) \cdot \Delta \overline P (t) + 
\frac{\sum\nolimits_{i = 1}^{dt} {P(t - dt + i)} }{dt} \quad ,
\end{equation}

\begin{equation}
\label{eq2}
\Delta \overline P (t) = \frac{\sum\nolimits_{i = 1}^{dt} {\{P(t - dt + i) 
\cdot (i - \frac{1 + dt}{2})\}} }{\sum\nolimits_{i = 1}^{dt} {(i - \frac{1 + 
dt}{2})^2} },
\end{equation}

\vspace{-3mm}
\noindent
where, dt is the number of past prices that the trader uses for this trend 
analysis.

We introduce the magnitude of price fluctuations by the absolute value of 
difference of the mean market price and the market price, $\vert \overline P 
(t) - P(t)\vert $, and assume that the market price is modeled by a kind of 
mean field type approximation that individual traders' action is replaced by 
random noises. The model is given by the following equation,

\begin{equation}
\label{eq3}
P(t + 1) = \overline P (t) + b_1 (t) \cdot \Delta \overline P (t) + b_2 
(t)\left| {\overline P (t) - P(t)} \right| + f(t),
\end{equation}
\vspace{-10mm}

\begin{center}
\begin{figure}
\begin{minipage}{.5\linewidth}
\centerline{\includegraphics[height=40mm]{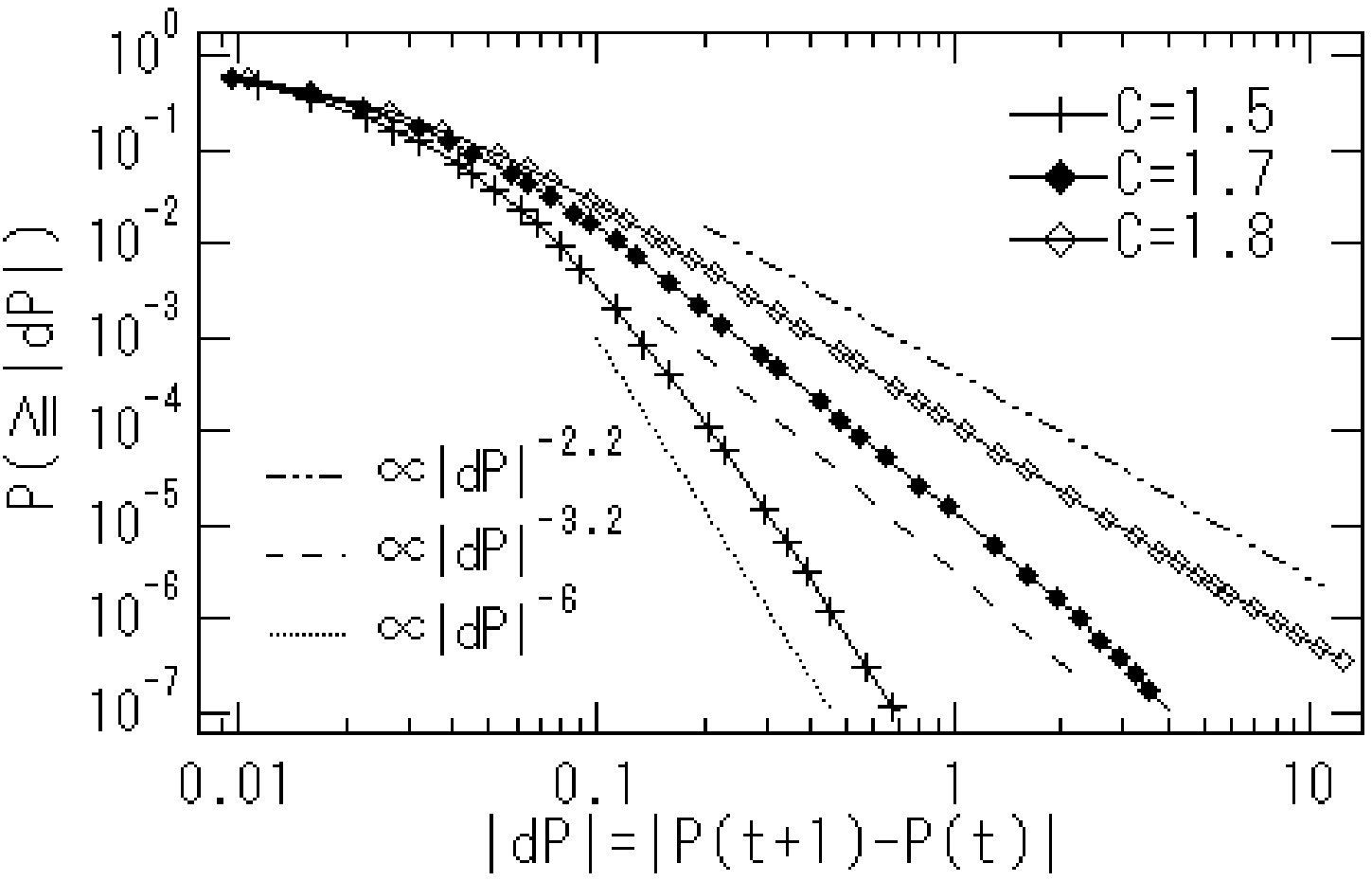}}
\label{fig3}
\caption[Fig.3 ]{Cumulative probability distributions of price changes of model with 
$dt = 50$ and $b_1 = 0$. The values of $C$ are 1.5, 1.7 and 1.8, 
respectively.}
\end{minipage}
\begin{minipage}{.5\linewidth}
\centerline{\includegraphics[height=40mm]{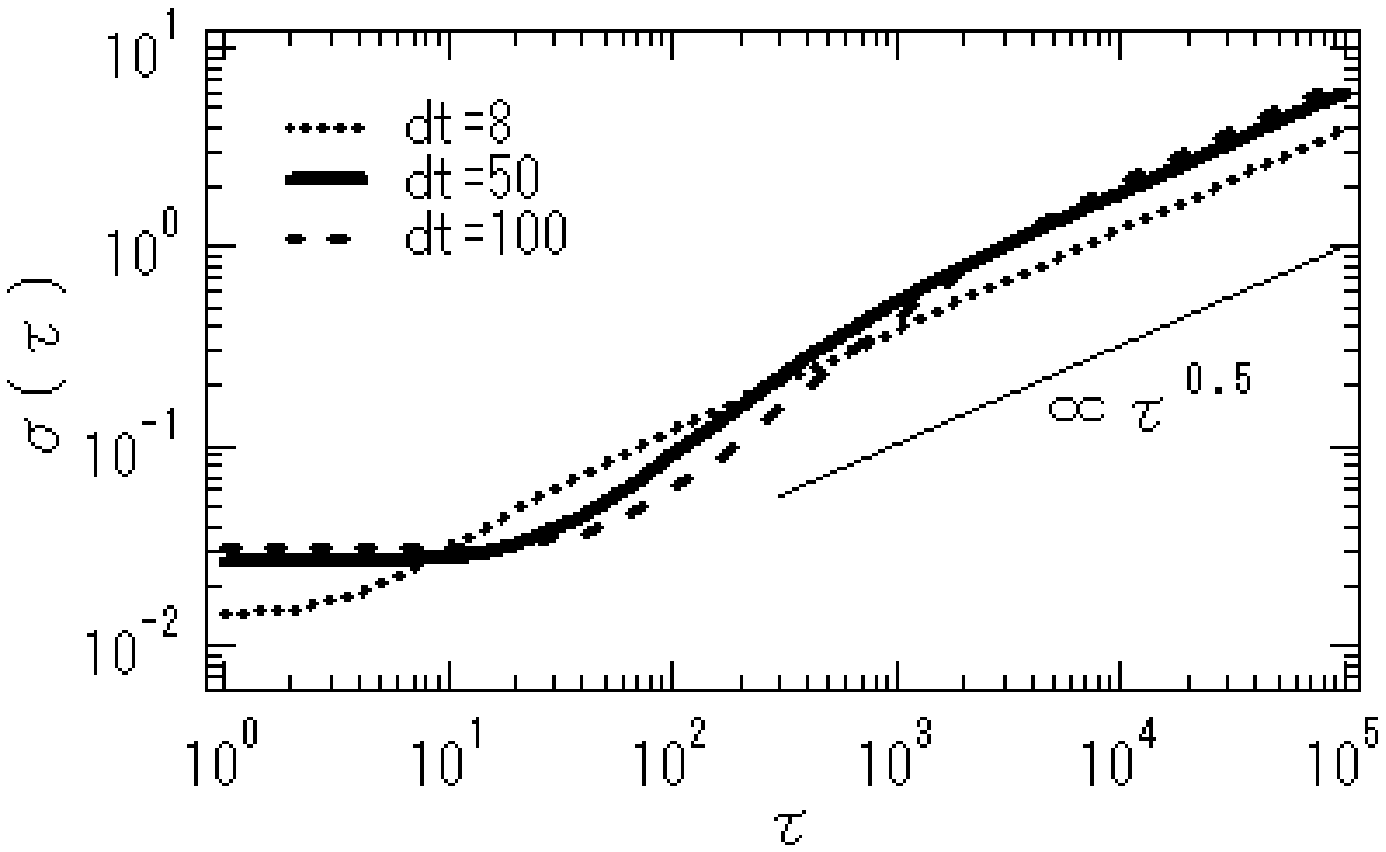}}
\label{fig4}
\caption[Fig.4 ]{The standard deviation of price diffusion of our model with $b_1 = 0$ 
and $C = 1.6$. The values of $dt$ are 8, 50, and 100.}
\end{minipage}
\end{figure}
\end{center}

\vspace{-3mm}
\begin{figure}[htbp]
\centerline{\includegraphics[height=38mm]{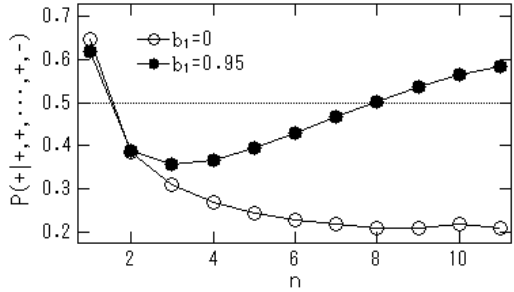}}
\label{fig5}
\caption[Fig.5 ]{Up-down statistics of our model. The probability of finding + 
(price-up) after $n$ successive +. ($\bullet$) and ($\circ$) denote the cases with $b_1 = 
0.95$ and $b_1 = 0$, respectively.}
\end{figure}

\noindent
where, $b_1 (t)$ denotes the coefficient of traders' response to the market 
trend at time $t$, $b_2 (t)$ is the coefficient of response to the magnitude 
of price fluctuations, and $f$ shows the white noise. In our numerical 
simulations $f$ is given by Gaussian noises with the average value 0 and the 
standard deviation 0.01. 

An example of actual yen-dollar rates and numerical simulation results are 
shown in Fig.2. It is intuitive confirmed that this model reproduces basic 
short time properties nicely.

\textbf{Effects of traders' basic strategy on price change} \\
Probability density of short time yen-dollar rate changes is known to have 
symmetric fat-tails approximated by a power law [2]. This basic property can 
be derived from our numerical model Eq.(\ref{eq3}) as shown in Fig.3. The parameters 
are $dt = 50$, $b_1 = 0$, and the value of $b_2 (t)$ is given randomly at 
each time by a uniform random number between $ - C$ and $ + C$, where $C$ is 
a given parameter. As confirmed in this figure the price change distribution 
is always approximated by a power law and its exponent depends on the value 
of $C$. This power law property is expected to be closely related to the 
multiplicative random process with additive noise [6]. The observed value of 
the exponent for the real data is about 3, so this result indicates that the 
best fit parameter of $C$ is about 1.7. 

 The non-standard diffusion characteristics of yen-dollar exchange rate 
fluctuation is described by the slow expansion of the standard deviation for 
time shorter than 100 ticks [1]. We calculate Eq.(\ref{eq3}) with different values 
of $dt$ and compared the diffusion characteristics as shown in Fig.4. It is 
found that the diffusion is slower for longer memory, and the empirical 
bending of this graph around 100 ticks is reproduced with a suitable choice 
of $dt$.

 It is known that up-down statistics of yen-dollar rates shows a quite 
non-trivial property that the conditional probability of successive same 
sign, P($+\vert +,+,\cdots $ \\$,+,-$), is smaller than 0.5 for $n < 8$ and 
larger than 0.5 for $n > 8$, where $n$ denotes the number of same signs[2]. In 
Fig.5 we can confirm that this property can be realized by our model when 
the parameter $b_1 $ is 0.95. Namely, the traders' trend-following action 
can be well modeled by Eq.(\ref{eq3}).

\textbf{Discussion} \\
We introduced a new type of market price equation that describes the short 
time market characteristics consistent with the real data. Our model is 
based on the assumption that the traders in open markets are generally 
predicting the future price using a moving average of latest prices. It is 
confirmed that this feedback of information is responsible for the slow 
diffusion, the power law distribution of price change and the 
trend-following action. In the case of yen-dollar market the best fit moving 
average is obtained with the time scale of the order of several minutes 
implying that traders are watching short time market behaviors only.




\end{document}